\documentclass[12pt]{iopart}
\usepackage[english]{babel}
\usepackage[latin1]{inputenc}
\usepackage{hyperref}

\usepackage{graphicx}
\graphicspath{{images_symm/}}
\usepackage{booktabs}
\usepackage{multirow}

\usepackage{amsmath}
\usepackage{xcolor}
\usepackage{mathrsfs}
\usepackage{iopams} 

\DeclareMathOperator{\Id}{ {\bf 1}}

\def\pmx{\begin{pmatrix}}
\def\emx{\end{pmatrix}}

\newcommand{\ket}[1]{|#1\rangle}
\newcommand{\bra}[1]{ \langle #1 \,  |}

\begin{document} 

\title{Quantum Hypergraph States}

\author{M. Rossi$^1$ $^\ast$, M. Huber$^{2,3,4}$, D. Bru{\ss}$^5$ and C. Macchiavello$^1$}

\address{$^1$ Dipartimento di Fisica and INFN-Sezione di Pavia,
Via Bassi 6, 27100 Pavia, Italy}
\address{$^2$ University of Bristol, Department of Mathematics, Bristol, BS8 1TW, U.K.}
\address{$^3$ ICFO-Institut de Ciencies Fotoniques, 08860 Castelldefels, Barcelona, Spain}
\address{$^4$ Universitat Autonoma de Barcelona, 08193 Bellaterra, Barcelona, Spain}
\address{$^5$Institut f{\"u}r Theoretische Physik III, 
Heinrich-Heine-Universit{\"a}t D{\"u}sseldorf, D-40225 D{\"u}sseldorf, Germany}

\ead{$^\ast$ matteo.rossi@unipv.it}

\begin{abstract}
We introduce a class of multiqubit quantum states which generalizes graph 
states. These states correspond to an underlying mathematical hypergraph, 
i.e. a graph where edges connecting more than two vertices are considered. We derive a generalised stabilizer formalism  to describe this class of states. 
We introduce the notion of $k$-uniformity and show that this gives rise to classes of states which are inequivalent under the action of the local Pauli group.
Finally we disclose a one-to-one correspondence with states employed in quantum 
algorithms, such as Deutsch-Jozsa's and Grover's.

\end{abstract}


\section{Introduction}

Quantum algorithms constitute one of the main applications of modern quantum information theory. They offer computational speed-up, that provably no classical system could ever exhibit \cite{nielsen}. The crucial quantum property for such a speed-up remains a heavily debated open question up to date (see e.g. \cite{Jozsa1,Gottesman,LindenPopescu,JozsaLinden,BrussMacchiavello, Matteospaper,VandenNest}). A famous way of implementing quantum algorithms is the measurement-based approach, where the computations are performed through the preparation of a highly entangled particular type of graph state (namely, a cluster state), which is subsequently processed via local measurements (introduced in Ref.~\cite{qc}).

Some of the most prominent algorithms, however, are often phrased in the circuit model (e.g. Grover's algorithm \cite{Grover} or Deutsch-Jozsa's algorithm \cite{DJ}), where the algorithm is usually formulated in terms of an initialization of a real equally weighted (REW) pure state (i.e. a superposition of all basis states with real amplitudes and equal probabilities), on which quantum gates subsequently act and a final measurement concludes the computation. Thus, this family of states plays a central role in several quantum algorithms. From the construction of graph states, as reviewed below, it is obvious that they are special cases of such REW states. Due to the special properties of graph states it can also easily be seen that in a many-body system they can be created using only particular two-body interactions. The first question we address in this work is whether all REW states  can be created using the two-body interactions occurring  in graph states. We show that this is not the case, i.e.  the set of REW states is strictly larger than the set of graph states.

From a physical point of view, where the $n$-qubits are a composite system of interacting spin 1/2 particles, it is then interesting to ask what kind of interactions are necessary to create all possible REW states. In this paper we answer this question by introducing hypergraph states, i.e. quantum states created by using up to $n$-body interactions of a given kind. We show that these states indeed cover 
all possible REW states, by providing an explicit simple procedure to find the associated hypergraph to a given REW state, and that they have an illustrative graph representation. We also find that they are stabilized by generalizations of the stabilizers of graph states, and that, by introducing the notion of $k$-uniformity, they can be shown to constitute a set of different entanglement classes under local Pauli operations.

The present work is structured as follows. In Sec. \ref{prel} we briefly review concepts related to standard graph states. We introduce and mathematically define $k$-uniform hypergraph states in Sec. \ref{inter}, and general hypergraph states in Sec. \ref{final}. In Sec. \ref{rew} the equivalence and connection with REW states employed in quantum algorithms is proven and discussed. We finally summarize our results  in Sec. \ref{conc} and discuss possible ways to extend our work.

\section{Standard graph states}
\label{prel}

\begin{figure}[t!]
\centering
\includegraphics[scale=.55]{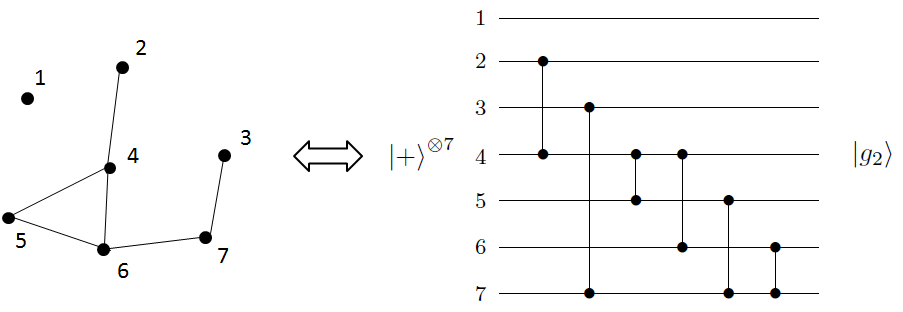}
\caption{Correspondence between a mathematical graph and the quantum state associated to it. Since controlled-$Z$ gates are symmetric, $C^2Z$ gates are here depicted as two dots connected by a vertical line. This notation will also be used for general controlled-$Z$ gates in the next figures. }\label{graph2}
\end{figure}

We will  here briefly review some basic concepts related to graph states 
(following Ref. \cite{vdn}), 
for fixing the notation and introducing concepts that will be useful later.
Given a mathematical graph $g_2=\{V,E\}$, i.e. a set of $n$ vertices 
$V$ and a set of edges $E$, one can find the corresponding quantum graph state as 
follows.
First, assign to each vertex a qubit and initialise each qubit as the state 
$\ket{+}=\frac{1}{\sqrt 2}(\ket{0}+\ket{1})$, so that the initial $n$-qubit 
state is given by $\ket{+}^{\otimes n}$. Then, perform controlled-$Z$ 
operations 
between any two qubits that are connected by an edge. By performing the operation 
$C^2Z_{i_1i_2}=diag(1,1,1,-1)$ for any two connected qubits $i_1$ and 
$i_2$, we get the corresponding quantum graph state
\begin{equation}
\ket{g_2}=\prod_{\{i_1,i_2\}\in E}C^2Z_{i_1i_2}\ket{+}^{\otimes n},
\label{graphstate}
\end{equation}
where $\{i_1,i_2\}\in E$ means that the two vertices $i_1$ and $i_2$ are connected by an edge.
This procedure is sketched for an example in  
Fig. \ref{graph2}, which clearly points out the correspondence.
In this work we will denote by $\Id, X, Y,$ and $Z$ the identity and the 
three Pauli matrices $\sigma_x, \sigma_y,$ and $\sigma_z$, respectively. 
We will also denote by $C^kZ_{i_1i_2...i_k}$ the general controlled-$Z$ gate 
acting on the $k$ qubits  labelled by $i_1i_2...i_k$. 
Notice that $k$ is an integer in the interval $1\leq k\leq n$, and by 
definition we take $C^1Z_{i_1}\equiv Z_{i_1}$. The gate 
$C^kZ_{i_1i_2...i_k}$ introduces a minus sign to the input state 
$\ket{11...1}_{i_1i_2...i_k}$, i.e. 
$C^kZ_{i_1i_2...i_k}\ket{11...1}_{i_1i_2...i_k}=-\ket{11...1}_{i_1i_2...i_k}$, 
and leaves all the other components of the computational basis unchanged. 
Hence the action of the controlled-$Z$ gate is invariant under permutations of
the $n$ qubits in the 
computational basis, thus any of the $k$ qubits on which $C^kZ_{i_1i_2...i_k}$ acts  
can be thought of as the target qubit. We will take $C^0Z\equiv -1$,
the reason will be clear later.

We call the set of all graph states $G_2$, and denote a general state taken from this set by $\ket{g_2}$. The subscript $2$ indicates that only two-body interactions, represented by $C^2Z$, are involved. It is easy to show, 
by counting all possible edge configurations, that the number of graph states 
is $2^{B(n,2)}$, where $B(n,2)$ is the binomial coefficient ``$n$ choose 
$2$''. 

Graph states can alternatively be defined by exploiting 
the stabilizer formalism: Given the graph $g_2$, one defines a set of 
operators $\{K_{i}^{(2)}\}$ for which the state $\ket{g_2}$ is a simultaneous eigenvector with 
eigenvalue one. Explicitly, for any vertex $i$ the correlation operation 
$K_{i}^{(2)}$ is given as follows:
\begin{equation}
K_{i}^{(2)}=X_i\otimes Z_{N(i)}=X_i\bigotimes_{j\in N(i)} Z_j,
\end{equation}
where $N(i)=\{j|\{i,j\}\in E\}$ is the neighbourhood of the vertex $i$, namely the vertices $j$ which are connected to $i$ by an edge. Again, the index 2 refers to a 2-body interaction, i.e. an edge
between 2 vertices.
Thus,
the set of $n$ operators $\{K_{i}^{(2)}\}_{i=1,2,...,n}$ uniquely defines the graph state $\ket{g_2}$ associated to the graph $g_2$, according to
\begin{equation}\label{stab}
K_{i}^{(2)}\ket{g_2}=\ket{g_2} \text{ for every } i=1,2,...,n.
\end{equation}
It can be shown that the set $\{K_i^{(2)}\}$  gives rise to a commutative subgroup called stabilizer (as each element of the group stabilizes the state $\ket{g_2}$, see Eq. \eqref{stab}) of the Pauli group on $n$ qubits, generated by the tensor product of the Pauli matrices $X$, $Y$ and $Z$.
The definitions of graph states based on the explicit procedure involving 
$C^2Z$ gates and on the stabilizer formalism can be shown to be 
equivalent \cite{vdn}.

\section{$k$-uniform hypergraph states}
\label{inter}

In this Section we generalise the notion of graph states allowing interactions which 
involve more than two parties. The mathematical tools needed to achieve this 
are $k$-uniform hypergraphs. A $k$-uniform hypergraph $g_k=\{V,E\}$ is 
a set of $n$ vertices $V$ with a set of edges $E$, where each edge connects exactly $k$ vertices,
and is called $k$-hyperedge (thus, a connected graph in the common sense 
is a 2-uniform hypergraph).

\begin{figure}[t!]
\centering
\includegraphics[scale=.55]{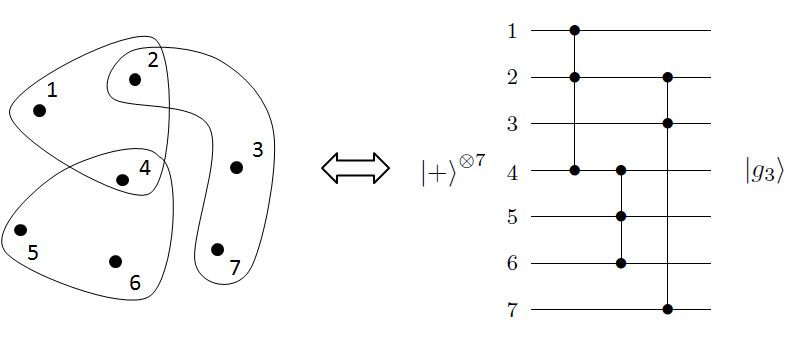}
\caption{Correspondence between a mathematical $3$-uniform hypergraph and the quantum state associated to it.
A hyperedge is visualised by a closed curve around a set of vertices.}\label{graph3}
\end{figure}

Given a $k$-uniform hypergraph, by following a similar procedure as before, 
one can find the corresponding $k$-uniform quantum hypergraph state as follows.
Assign to each vertex a qubit and initialise each qubit in the state 
$\ket{+}$. 
Wherever there is a $k$-hyperedge, perform a controlled-$Z$ operation between the $k$ connected qubits. Formally, if the qubits $i_1, i_2, ..., i_k$  are connected, then perform the operation $C^kZ_{i_1i_2...i_k}$. In this way 
we arrive at the state
\begin{equation}\label{khyperdef}
\ket{g_k}=\prod_{\{i_1,i_2,...,i_k\}\in E}C^kZ_{i_1i_2...i_k}\ket{+}^{\otimes n},
\end{equation}
where $\{i_1,i_2,....,i_k\}\in E$ means that the $k$ vertices are connected by a $k$-hyperedge.
In Fig. \ref{graph3} we show an explicit example of 
 a 3-uniform hypergraph and the  circuit implementation 
of the corresponding quantum hypergraph state.

For a fixed $k$ with $1\leq k\leq n$, we call the class of $k$-uniform hypergraph states  $G_k$ and 
denote the quantum state associated with the general $k$-uniform hypergraph $g_k$ as $\ket{g_k}$. The case $k=1$ can be simply thought of as $Z$ gates acting 
locally on single qubits, while the case $k=n$ is the only one involving 
interactions among all $n$ qubits, namely $C^nZ_{i_1i_2...i_n}$. Obviously, by setting $k=2$ we recover the class of graph states. By the same counting argument used above, for fixed $k$ the number of possible $k$-uniform hypergraphs 
is given by $2^{B(n,k)}$.
We will now show that $k$-uniform hypergraph states can be described in terms 
of a generalised stabilizer formalism.
Given a $k$-uniform hypergraph $g_k$, for each vertex $i=1,2,...,n$ we define the correlation operator
\begin{equation}\label{kstab}
K_{i}^{(k)}=X_{i}\otimes C^{k-1}Z_{N(i)}
=X_i\otimes \prod_{(i_1,i_2,...,i_{k-1})\in N(i)} C^{k-1}Z_{i_1i_2...i_{k-1}},
\end{equation}
where the neighbourhood $N(i)$ of the vertex $i$ is given by $N(i)=\{(i_1,i_2,...,i_{k-1})|\\ \{i,i_1,i_2,...,i_{k-1}\}\in E\}$, namely all $k-1$-tuples $(i_1,i_2,...,i_{k-1})$ of vertices connected to $i$ via a $k$-hyperedge. Notice that, if $k$-body interactions are involved, then the stabilizers are defined in terms of controlled-$Z$ gates acting on $k-1$ qubits. 
Hence, the generalized stabilizers for general $k$ no longer belong to the Pauli group, except in the case of graph states where we recover the stabilizer operators given by local Pauli matrices. These operators nevertheless can be shown to ``stabilize" the regarded state as follows.

The $k$-uniform hypergraph state $\ket{g_k}$  corresponding to $g_k$ is then defined as the unique eigenvector with  eigenvalues one of the $n$ operators $\{K_{i}^{(k)}\}$, namely
\begin{equation}
K_{i}^{(k)}\ket{g_k}=\ket{g_k} \text{ for every } i=1,2,...,n.
\end{equation}
The set of the operators generated by $\{K_{i}^{(k)}\}_{i=1,2,...,n}$ is an Abelian 
group (see Appendix B). This Abelian group can be thought of as a subgroup of a 
generalized Pauli group which contains, besides the tensor product of 
Pauli matrices, also $C^{k-1}Z$ gates acting on any $k-1$-tuple of qubits 
as generators. 
Furthermore, as for standard graph states, the definition following the generalised stabilizer is completely equivalent to the constructive procedure involving controlled-$C^kZ$ operations. The equivalence can be
explicitly derived in the more general case of non-uniform hypergraphs, that will be considered in the following, of which the
 $k$-uniform hypergraphs are  a strict subset.

The classification induced by $k$-uniformity allows us to prove that two sets 
$G_k$ and $G_{k'}$ cannot be connected by local Pauli operators 
for $k\neq k'$ (apart from the 
trivial separable state $\ket{+}^{\otimes n}$ which corresponds to the empty graph and thus is already contained in every class). Therefore, each set $G_k$ gives rise to an inequivalent class under the action of the local Pauli group of $n$ qubits (see Appendix A). It is, however, an open
question whether two sets $G_k$ and $G_{k'}$ with
$k\neq k'$ are inequivalent under the action
of general local unitaries. An affirmative answer
to this question would imply a corresponding
multipartite entanglement classification.
Notice nevertheless that for the class of standard graph states, i.e. $G_2$, it is known that there exist states which are local unitary equivalent, but not local Clifford equivalent \cite{lc}. This finally suggests that the local unitary inequivalence of $G_k$ and $G_{k'}$ might be a hard problem to solve.

\section{Hypergraph states}
\label{final}

We  are now ready to define a general hypergraph state as follows. 
A hypergraph $g_{\leq n}=\{V,E\}$ is a set of $n$ vertices $V$ with a set of  
hyperedges $E$ of any order $k$ (thus $k$ is no longer fixed but may
range from $1$ to $n$).
Given a mathematical hypergraph, the corresponding quantum state can be found by following the three steps:
Assign to each vertex a qubit and initialise each qubit as $\ket{+}$ 
(the total initial state is then given by $\ket{+}^{\otimes n}$).
Wherever there is a hyperedge, perform a controlled-$Z$ operation between 
all  connected qubits. Formally, if the qubits $i_1, i_2, ..., i_k$  are connected by a $k$-hyperedge, then perform the operation $C^kZ_{i_1i_2...i_k}$. So eventually we get the quantum state
\begin{equation}
\ket{g_{\leq n}}=\prod_{k=1}^n\prod_{\{i_1,i_2,...,i_k\}\in E}C^kZ_{i_1i_2...i_k}\ket{+}^{\otimes n},
\end{equation}
where $\{i_1,i_2,....,i_k\}\in E$ means that 
the $k$ vertices are connected by a $k$-hyperedge. Notice that the product of $k=1,2,...,n$ accounts for different types of hyperedges in the hypergraph.

\begin{figure}[t!]
\centering
\includegraphics[scale=.55]{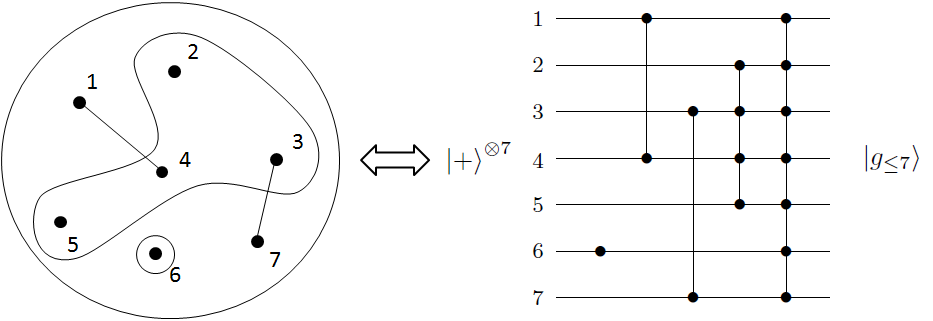}
\caption{Correspondence between a mathematical hypergraph and the quantum 
hypergraph state. The circle around vertex $6$ stands for a local $Z$ gate, corresponding to a hyperedge of order $k=1$, while the big circle around all vertices corresponds to a full-body interaction.}\label{graphn}
\end{figure}

We illustrate the correspondence with an example in Fig. \ref{graphn}. There, some hyperedges connecting $1,2,4$ and $7$ vertices appear, and thus controlled-$Z$ operations acting on $1,2,4$ and $7$ qubits must be considered.

We denote 
the set of all general hypergraph states for graphs with $n$ vertices
 as $G_{\leq n}$, 
indicating that hyperedges connecting up to $n$ vertices are present. An
element of this set, corresponding to the particular graph  $g_{\leq n}$,
is called $\ket{g_{\leq n}}$. 
 Each set of 
$k$-uniform hypergraphs, for fixed $k$, is obviously a subset of all possible 
hypergraphs.
In order to count the number of hypergraph states, we exploit any possible combination of $k$-uniform hypergraphs. Since the number of the latter ones for 
fixed $k$ is given by $2^{B(n,k)}$, the total number of hypergraph states for $n$ vertices turns out to be $\prod_{k=1}^n2^{B(n,k)}=2^{\sum_{k=1}^n B(n,k)}=2^{2^n-1}$.

We will  now describe general hypergraph states in {a generalised} stabilizer formalism. 
Given a  general hypergraph, for any vertex $i$ we define the following correlation operator
\begin{equation}\label{generators}
K_{i}=X_{i}\otimes\prod_{k=1}^{n} C^{k-1}Z_{N(i)} 
=X_i\otimes\prod_{k=1}^{n}\prod_{(i_1,i_2,...,i_{k-1})\in N(i)} C^{k-1}Z_{i_1i_2...i_{k-1}},
\end{equation}
where the product over $k$ takes into account all kinds of hyperedges that appear. For any value of $k$, the neighbourhood $N(i)$ of the vertex $i$ is still defined as $N(i)=\{(i_1,i_2,...,i_{k-1})|\{i,i_1,i_2,...,i_{k-1}\}\in E\}$. Different kinds of neighbourhoods can obviously appear in this scenario (single vertices, couples and in general $k-1$-tuples), depending on the order $k$ of the hyperedges that connect
the vertex $i$ to other vertices. For instance, in Fig. \ref{graphn} the neighbourhood $N(4)$ of vertex $4$ 
consists of the tuples  $1$, $(2,3,5)$ and $(1,2,3,5,6,7)$, since the vertex is connected via 
the depicted hyperedges of order $2$, $4$ and $7$. 
We introduce a generalised stabilizer group, being generated by the set $\{K_i\}$, which stabilizes the corresponding hypergraph state. We show  in Appendix B that this group is Abelian. The unique hypergraph state corresponding to the set $\{K_{i}\}$ is then defined as the unique eigenvector with eigenvalues one of any generator $K_{i}$, i.e.
\begin{equation}
K_i\ket{g_{\leq n}}=\ket{g_{\leq n}} \text{ for every } i=1,2,...,n.
\end{equation}
Furthermore, one can show (see Appendix C) that the definition according to the {generalised} stabilizer formalism is equivalent to the one given above in terms of controlled-$Z$ gates.

\section{REW states and their equivalence with hypergraph states}
\label{rew}

Let us now introduce a different set $G_\pm$ of $n$-qubit states, namely the ``real equally weighted" (REW) pure states, defined as
\begin{equation}
\ket{f}=\frac{1}{2^{n/2}}\sum_{x=0}^{2^n-1}(-1)^{f(x)}\ket{x},
\end{equation}
where $\ket{x}$ are the computational basis states, while $f(x)$ is a Boolean function, i.e. $f:\{0,1\}^n\rightarrow\{0,1\}$. The state $\ket{f}$ is uniquely defined by the function $f$ via the signs (either plus or minus) in front of each component $x$ of the computational basis. According to this we can count the number of 
REW states, which turns out to be $2^{2^n-1}$. 
These states are employed in many quantum protocols, and in particular in the
well-known quantum algorithms of Deutsch-Jozsa and Grover.
Notice that a more general class of equally weighted states, with generic phase
factors in front of each computational basis state, and an 
explicit method to generate them was analysed in \cite{kraus}.

Is there any relation between REW states and graph states or hypergraph
states? From the construction in Eq. (\ref{graphstate}) it is clear that 
all graph states are REW states, since the action of $C^2Z$ can only produce some 
minus signs. Thus,  $G_2\subseteq G_\pm$ holds. But is the reverse also true,
i.e. are all REW states graph states? 
 A first hint that this is not the case comes from the 
fact that the number of REW states is exponentially larger  than the number
of graph states, i.e. 
$2^{2^n-1}$ versus $2^{B(n,2)}$, see above. In order to prove that not every
REW state is a graph state,
we provide a counterexample, given by the state
\begin{equation}\label{grover}
\ket{f}=\frac{1}{\sqrt{8}}(\ket{000}+ \ket{001}+\ket{010}+\ket{011}
+ \ket{100}+\ket{101}+\ket{110}-\ket{111}).
\end{equation}
It is easy to show that the geometric measure of genuine multipartite entanglement \cite{wg} of the state above is $E_2(\ket{f})=1/4$ 
\cite{Matteospaper}, however every connected graph state has a multipartite geometric 
measure $E_2(\ket{g_2})\geq 1/2$ \cite{guehne}. 

Notice that, by construction, graph states involve only particular two-body interactions, 
which are not sufficient to achieve all REW states. 
We will now investigate the relation between REW states and 
the wider class of hypergraph states and state our main 
result:
{\em The set $G_{\pm}$ of REW states and the set $G_{\leq n}$ of hypergraph states 
coincide.} We prove this statement as follows:
The inclusion $G_{\leq n}\subseteq G_{\pm}$ is trivial, since any 
$\ket{g_{\leq n}}$ is obtained from $\ket{+}^{\otimes n}$ by applying 
controlled-$Z$ gates. The opposite inclusion $G_{\pm}\subseteq G_{\leq n}$ 
can be proved by the following constructive approach. Suppose we are given a 
REW state 
$\ket{f}$, then the following procedure leads to the underlying hypergraph.
First, erase all the minus signs of the states with one excitation, i.e. of 
the form $\ket{0...01_j0...0}$, by applying local $Z_j$ gates. Notice that, 
by doing this, we might create unnecessary minus signs in front of states 
with more than one excitation.
Second, apply $C^2Z$ gates in order to erase the negative signs in front of 
the components with two excitations (either coming from  the original state 
$\ket{f}$ or as by-products of the previous step). Observe that, since $C^2Z$ 
acts non-trivially only on states with more than one excitation, the minus 
signs previously erased will remain untouched.
As a general rule, apply $C^kZ$ operations, from $k=1$ until $k=n$, 
erasing successively
the minus signs in front of the components of the computational basis. 
In general, at the step $k$, we have erased the minus signs in 
front of the states with up to $k$ excitations.
The set of gates that are needed to transform $\ket{f}$ back to 
$\ket{+}^{\otimes n}$ provides the underlying hypergraph for the  
REW state under examination.
Notice that, since the procedure is uniquely defined according to the REW state from which we start, the underlying hypergraph is unique. Therefore, the correspondence between the sets $G_\pm$ and $G_{\leq n}$ is one-to-one.

\begin{figure}[t!]
\centering
\includegraphics[scale=.6]{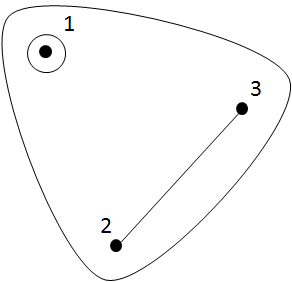}
\caption{Hypergraph corresponding to the REW state given by Eq. \eqref{pm}.}\label{under}
\end{figure}

As an explicit example consider the three-qubit REW state 
\begin{equation}\label{pm}
\ket{f}=\frac{1}{\sqrt{8}}(\ket{000}+ \ket{001}+\ket{010}-\ket{011}\\
- \ket{100}-\ket{101}-\ket{110}-\ket{111}).
\end{equation}
It is straightforward to see that the sequence of transformations 
$Z_1, C^2Z_{23}, C^3Z_{123}$ applied to the above state
leads to the initial state $\ket{+}^{\otimes 3}$.
Therefore the hypergraph corresponding to \eqref{pm} is the one depicted
in Fig. \ref{under}. 

Other interesting examples are REW states employed in quantum algorithms. 
In Grover's algorithm for instance, REW states with only one minus sign 
appear, such as the state \eqref{grover} for three qubits (the minus 
sign marks the single solution of the search problem). 
It is easy to see that, when the number of minus signs is odd, the 
REW state always involves a controlled-$Z$ gate acting on all the 
qubits, therefore involving $n$-body interactions.
On the contrary, for REW states employed in Deutsch-Jozsa's algorithm, such a 
gate is never needed, since the function $f$ is either constant or balanced 
(balanced means that the number of minus signs equals the number of plus 
signs), while the application of a controlled $Z$ gate acting on all qubits
would change just one sign in the $n$-qubit state, therefore necessarily 
leading to an unbalanced function.
An explicit example of a balanced state is given by modifying the state in (\ref{pm}) such that a plus sign is in front of the component $\ket{111}$. Such a state would be
generated by a sequence of the $Z_1$ and the $C^2Z_{23}$ gates, without
the application of $C^3Z_{123}$.

\section{Conclusions}
\label{conc}

In conclusion, we have introduced the class of 
quantum hypergraph states, which are associated to corresponding mathematical hypergraphs and are stabilized by nonlocal observables. 
We introduced the notion of $k$-uniformity and proved that this gives rise to classes of states which are inequivalent under the action of the local Pauli group.
We showed that there is a one-to-one  correspondence between the set of hypergraph states  and the set of real equally weighted states, which are  essential for quantum algorithms. A constructive method was introduced which allows us to generate the hypergraph  underlying a given real equally weighted state,  i.e. a quantum state encoding a given Boolean function $f$.
We have discussed the types of many-body interactions needed to generate general
hypergraph states in a Hamiltonian description, going beyond the two-body interaction that characterises graph states.

For future studies, since the class of hypergraph states naturally generalises the class of graph states, it will be of great interest to ask whether some of the 
many results about the latter, such as for instance measurement-based quantum computing \cite{qc}, entanglement witnessing \cite{guehne}, and quantum error correcting techniques \cite{gott2, schling}, can be extended to the former. Some achievement in this sense already exists, mainly related to purification protocols \cite{barbara}. Furthermore, this larger class of states
may enable even more applications and quantum protocols, especially in connection to already existing algorithms employing hypergraphs, as e.g. the 3-SAT problem \cite{3sat}.

While finishing this manuscript we learnt about related work \cite{Qu}  
which contains a similar analysis. Subsequent works by some of the same authors address the issues of the characterization of three-qubit hypergraph states \cite{china1}, and the relationship among hypergraph states, locally maximally entangleable states and $W$ states \cite{china2}.

\section*{Acknowledgements}

We would like to thank all the Quantum Information Theory Group at the HHU D\"usseldorf, in particular Hermann Kampermann, for stimulating discussions. We also thank Barbara Kraus, Marti Cuquet, and Andreas Winter for discussions about local unitary equivalence. MR gratefully acknowledges support from DAAD and fruitful discussions with Martin Hofmann. MH acknowledges funding from the FP7-MarieCurie grant ``Quacocos" and the hospitality of D\"usseldorf. This work was financially supported by DFG.

\section*{Appendix A: inequivalence of $k$-uniform hypergraph states under the local Pauli group}

In this Appendix we prove that every $k$-uniform hypergraph state cannot be transformed to any other $k'$-uniform hypergraph with $k\neq k'$, by the only action of local Pauli operators, namely $X$, $Y$ and $Z$.

Let us rewrite a general hypergraph state in the more convenient form
\begin{equation}
|g_{\leq n}\rangle=\frac{1}{2^{n/2}}\sum_{x=0}^{2^n-1} c_{\alpha^x_l}|x\rangle,
\end{equation}
where $\alpha^x_l$ denotes the set of cardinality $l$ of subsystems of the state $x$ that are in the state $|1\rangle$. In other words, given the state $\ket{x}$, $\alpha^x_l$ represents the set of indices corresponding to qubits where the excitations are. Then, having in mind that a general $k$-uniform hypergraph can be created from  $|+\rangle^{\otimes n}$ using $\prod_j{C^kZ_{\alpha^j_k}}$ ($\alpha^j_k$ are index sets of cardinality $k$ referring to the vertices on which the controlled $Z$ operations act, for a given hyperedge $j$), it is easy to see that for any $k$-uniform state there is at least one $c_{\alpha_k}$ negative (condition C1) and all $c_{\alpha_{k'}}$ with $k'<k$ are positive (condition C2).

In the following we prove that, starting from a $k'<k$-uniform hypergraph state, it is not possible to transform it into a $k$-uniform one by only using local $X$ and $Z$. Notice that, as $Y=iXZ$, the $Y$ operations are already considered. Furthermore, since $X$ and $Z$ anticommute, it is not restrictive to apply always $Z$ before $X$. As a result the two following cases describe the most general strategy we could apply.

\textit{Case 1)} We just use any $k'<k$ controlled $Z$ operations (which includes local $Z$'s when $k'=1$). This nevertheless fails always because to make $c_{\alpha_k}$ negative you generate at least one $c_{\alpha_{k'}}$ with $k'<k$ 
which is negative as well which contradicts C2.

\textit{Case 2)} We apply arbitrary $k'<k$ controlled $Z$ operations and then include any number of $X$ gates anywhere. 
We will now show that this  procedure will fail again. Let us denote as $c_{\beta_l}$ the coefficient that will afterwards be transformed to
the negative coefficient $c_{\alpha_k}$ ($l$ is of course arbitrary). We then need to apply $X$ in $\gamma_{l'}\equiv(\alpha_k\cup\beta_l)\setminus(\alpha_k\cap\beta_l)$ (such that $c_{\beta_l}\rightarrow c_{\alpha_k}$ and C1 holds). Now, since the action of $X$'s cannot change the sign of the coefficient $c_{\beta_l}$, the number of $C^{k'}Z_{\alpha_{k'}}$ operations we apply in the set $\beta_l$ must be odd (thus the number of different subsets $\alpha_{k'}$ must be odd as well). Let us denote this number as $N_{\beta_l}$, and in the following $N_{S}$ will always denote the number of sets $\alpha_{k'}$ (coming from $C^{k'}Z_{\alpha_{k'}}$ operations) included in the general set of indices $S$. We can then distinguish four different cases that may happen, summarised in Table \ref{tabla}.
\begin{table}[h!]
\centering
\renewcommand{\arraystretch}{1.1}
\begin{tabular}{c|c|c|c|c}\hline
$N_{\beta_l\setminus\alpha_k}$ & $N_{cr}$ & $N_{\beta_l\cap\alpha_k}$ & $N_{\alpha_k\setminus\beta_l}$ & $N_{\gamma_{l'}}$ \\ \hline
\multirow{2}*{odd} & \multirow{2}*{odd} & \multirow{2}*{odd} & even & odd $(1)$\\ 
 & & & odd & even $(2)$\\ \hline
\multirow{2}*{even} & \multirow{2}*{even} & \multirow{2}*{odd} & even & even $(3)$\\ 
 & & & odd & odd $(4)$\\ \hline
 \multirow{2}*{even} & \multirow{2}*{odd} & \multirow{2}*{even} & even & even $(5)$\\ 
 & & & odd & odd $(6)$\\ \hline
 \multirow{2}*{odd} & \multirow{2}*{even} & \multirow{2}*{even} & even & odd $(7)$\\ 
 & & & odd & even $(8)$\\ \hline
\end{tabular}
\caption{All possible cases for index sets - for an explanation, see main text.}\label{tabla}
\end{table}

\begin{figure}[h!]
\centering
\includegraphics[scale=.5]{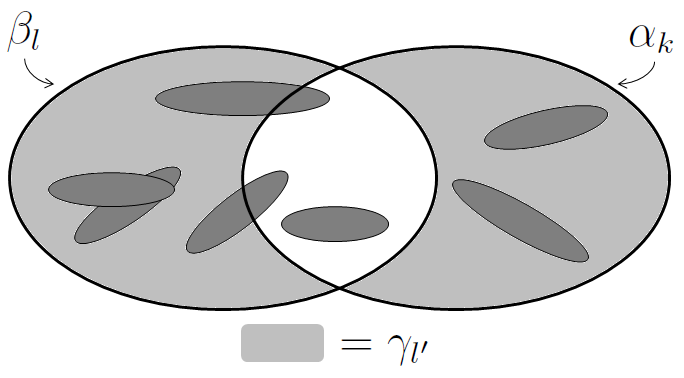}
\caption{Drawing showing an example for possible index sets. Each dark grey circle represents a set $\alpha_{k'}$. In this case $N_{\beta_l\setminus\alpha_k}=2$, $N_{cr}=2$, $N_{\beta_l\cap\alpha_k}=1$, $N_{\alpha_k\setminus\beta_l}=2$ and $N_{\gamma_{l'}}=4$. This is an example of case $(3)$ in Table \ref{tabla}. Notice that we do not take into account the case where subsets $\alpha_{k'}$ cross the border between the set $\alpha_k\setminus\beta_l$ and the intersection, since it is easy to see that this case never affects our counting.}
\label{set}
\end{figure}

By $N_{cr}$ we mean the subsets $\alpha_{k'}$ that lie across the border of the set $\beta_l\setminus\alpha_k$ and the intersection $\beta_l\cap\alpha_k$ (see Fig. \ref{set} for a comprehensible explanation). Notice that $N_{\beta_l}=N_{\beta_l\setminus\alpha_k}+N_{cr}+N_{\beta_l\cap\alpha_k}$ must be odd from the hypothesis, while the number of sets $\alpha_{k'}$ in $\alpha_k\setminus\beta_l$, namely $N_{\alpha_k\setminus\beta_l}$, is instead not determined, and might be either odd or even. Notice that the number of subsets $\alpha_{k'}$  in $\gamma_{l'}$ is given by $N_{\gamma_{l'}}=N_{\beta_l\setminus\alpha_k}+N_{\alpha_k\setminus\beta_l}$.  

For the cases $(1)-(4)-(6)-(7)$ the contradiction to C2 can be found by realising that $c_{\gamma_{l'}}=-1$ (since $N_{\gamma_{l'}}$ is odd). This coefficient will be mapped into $c_{\{\}}=-1$ (the coefficient of the state with all zeros) by the action of $X_{\gamma_{l'}}$, and thus showing a contradiction to C2.

For the cases $(2)-(3)$ the contradiction to C2 is given by $c_{\gamma_{l'}\cup(\alpha_{k'}\in\beta_l\cap\alpha_k)}=-1$ (since $N_{\gamma_{l'}}+N_{\beta_l\cap\alpha_k}$ is odd), which becomes $c_{(\alpha_{k'}\in\beta_l\cap\alpha_k)}=-1$ after $X_{\gamma_{l'}}$. By ${\gamma_{l'}\cup(\alpha_{k'}\in\beta_l\cap\alpha_k)}$ we mean the union between the  set $\gamma_{l'}$ and the sets $\alpha_{k'}$ which belong to the intersection $\beta_l\cap\alpha_k$.

For the case $(8)$ the contradiction to C2 is $c_{\gamma_{l'}\setminus (\alpha_{k'}\in\alpha_k\setminus\beta_l)}=-1$, since this coefficient is mapped by $X_{\gamma_{l'}}$ into $c_{(\alpha_{k'}\in\alpha_k\setminus\beta_l)}=-1$. By $\gamma_{l'}\setminus(\alpha_{k'}\in\alpha_k\setminus\beta_l)$ we mean the difference between the  set $\gamma_{l'}$ and the sets $\alpha_{k'}$ which belong to the set given by $\alpha_k\setminus\beta_l$.

Regarding the case $(5)$,
since $N_{cr}$ is odd we can always find a subset $\theta_t$ in the 
intersection ${\beta_l\cap\alpha_k}$ with cardinality $t<k$ such that the coefficient $c_{(\beta_l\setminus\alpha_k)\cup\theta_t}=-1$. Therefore, when we apply $X_{\gamma_{l'}}$ this will be mapped into $c_{(\alpha_k\setminus\beta_l)\cup\theta_t}=-1$ which clearly shows a contradiction to C2 since $(\alpha_k\setminus\beta_l)\cup\theta_t$ is a subset of $\alpha_k$ with cardinality strictly smaller than $k$.

\section*{Appendix B: group structure of the generalised stabilizer operators}

We now prove that the operators $\{K_i\}_{i=1,2,...,n}$ defined in Eq. \eqref{generators} generate an Abelian group. The group properties follow immediately: the closure is given by construction, the associativity by the matrix algebra, the identity and the inverse belong to the set since $K_i^2=\Id$ and $K_i=K_i^\dagger$ hold, respectively.

On the other hand, the commutativity can be proved as follows. Suppose we are given $K_i$ and $K_j$ with $i\neq j$, otherwise everything trivialises. Since the concept of neighbourhood is symmetric we can keep $K_i$ fixed and see what happens for different $K_j$. If $j$ is not in $N(i)$ then the stabilizer operators trivially commute. Therefore, the only situations we have to check is when $j\in N(i)$, namely when some of the $CZ$ gates acting on $N(i)$ in the definition of $K_i$ involves also the qubit $j$. Each of these gates takes the form $C^{k}Z_{ji_1i_2...i_{k-1}}$ (with $k$ arbitrary) and generally does not commute with $X_j$ defining $K_j$.

It is then easy to see that, in order to prove that $[K_i,K_j]=0$, it is sufficient to show that
\begin{equation}\label{comm}
[(X_i\otimes C^{k}Z_{ji_1i_2...i_{k-1}}),(C^{k}Z_{ii_1i_2...i_{k-1}}\otimes X_j)]=0,
\end{equation}
for any number of qubits $k-1$ and vertices $i_1i_2...i_{k-1}$. This is because we can think to commute the two operators $K_i$ and $K_j$ by following a step-by-step procedure consisting in swapping each term $(X_i\otimes C^{k}Z_{ji_1i_2...i_{k-1}})$ of $K_i$ with the corresponding term $(C^{k}Z_{ii_1i_2...i_{k-1}}\otimes X_j)$ of $K_j$. 

In order to prove Eq. \eqref{comm}, we rewrite the general controlled $Z$ gate acting on $k$ qubits as
\begin{equation}\label{decomp}
C^{k}Z_{ji_1i_2...i_{k-1}}=\Id_j\otimes(\Id-P)_{i_1i_2...i_{k-1}}+Z_j\otimes P_{i_1i_2...i_{k-1}},
\end{equation}
where $P_{i_1i_2...i_{k-1}}=\ket{11...1}_{i_1i_2...i_{k-1}}\bra{11...1}$. Then, by exploiting the anti-commutativity of Pauli matrices, it follows that
\begin{eqnarray}
(X_i\otimes &C^{k}Z_{ji_1i_2...i_{k-1}})(C^{k}Z_{ii_1i_2...i_{k-1}}\otimes X_j)
=\nonumber \\
&=(X_i\otimes\Id_j\otimes(\Id-P)_{i_1i_2...i_{k-1}}+X_i\otimes Z_j\otimes  P_{i_1i_2...i_{k-1}})\nonumber\\
&  \times (\Id_i\otimes X_j\otimes(\Id-P)_{i_1i_2...i_{k-1}}+Z_i\otimes X_j\otimes P_{i_1i_2...i_{k-1}})\nonumber \\
&=X_i\otimes X_j\otimes(\Id-P)_{i_1i_2...i_{k-1}}+X_iZ_i\otimes Z_jX_j\otimes P_{i_1i_2...i_{k-1}}\nonumber \\
&=X_i\otimes X_j\otimes(\Id-P)_{i_1i_2...i_{k-1}}+Z_iX_i\otimes X_jZ_j\otimes P_{i_1i_2...i_{k-1}}\nonumber\\
&=(C^{k}Z_{ii_1i_2...i_{k-1}}\otimes X_j)(X_i\otimes C^{k}Z_{ji_1i_2...i_{k-1}}).
\end{eqnarray}
Thus, since the commutativity relation stated in Eq. \eqref{comm} holds for any $k-1$ and qubits $i_1i_2...i_{k-1}$, the commutativity of any two stabilizers defined by \eqref{generators} finally follows.

\section*{Appendix C: equivalence of the circuital definition and the stabilizers description}

In order to prove that the two definitions stated in the main article are equivalent we will essentially follow Ref. \cite{vdn}. The proof is by induction on the number of hyperedges. The case with no hyperedges is trivially stabilized by the Pauli matrices $\{X_1,X_2,..., X_n\}$, since the corresponding graph state is given by $\ket{+}^{\otimes n}$. Suppose now a general hypergraph state $\ket{g_{\leq n}}$, corresponding to the hypergraph $g_{\leq n}$, is stabilized by $K_i$ as defined in \eqref{generators}, namely $K_i\ket{g_{\leq n}}=\ket{g_{\leq n}}$. We want to show that if we apply $C^kZ_{i_1i_2...i_k}$ to $\ket{g_{\leq n}}$, the new hypergraph state  $\ket{g'_{\leq n}}=C^kZ_{i_1i_2...i_k}\ket{g_{\leq n}}$ is stabilized by a new stabilizer generated by $K'_i$, derived from the hypergraph $g'_{\leq n}$ where the $k$-hyperedge $\{i_1,i_2,...i_k,\}$ is added (or removed). Namely we want to prove that $K'_i\ket{g'_{\leq n}}=\ket{g'_{\leq n}}$, where $K'_i$ is defined according to \eqref{generators} for the new hypergraph $g'_{\leq n}$.  If we consider $i\neq i_1,i_2,...,i_k$ then by definition we have $K_i'=K_i$ and, since $[K_{i},C^kZ_{i_1i_2...i_k}]=0$, the following holds
\begin{equation}
K'_{i}\ket{g'_{\leq n}}=\ket{g'_{\leq n}} \text{ for }i\neq i_1,i_2,...,i_k.
\end{equation}
So, as for the proof regarding the commutativity of the stabilizer group, we need to focus only on the operators $\{K'_{i_1},K'_{i_2},...,K'_{i_k}\}$, since the others are not affected by the action of $C^kZ_{i_1i_2...i_k}$. Keeping in mind the decomposition \eqref{decomp} of $C^kZ_{i_1i_2...i_k}$, it is then easy to show that for every $i=i_1,i_2,...,i_k$ the following relation holds
\begin{equation}
C^kZ_{i_1i_2...i_k}K_{i}C^kZ_{i_1i_2...i_k}=C^{k-1}Z_{i_2...i_k}K_{i}=K'_i\text{ for } i=i_1,i_2,...,i_k.
\end{equation}
Therefore, by exploiting the equation above, we can easily show that the hypergraph state $\ket{g'_{\leq n}}$ is eigenstate of $K'_i$ with eigenvalue one for vertices $i=i_1,i_2,...,i_k$. Hence, it follows that the hypergraph state $\ket{g'_{\leq n}}$ is stabilized by any $K'_i$ with $i=i_1,i_2,...,i_n$, which are the correlation operators that can be defined according to the hypergraph $g'_{\leq n}$.

\end{document}